\documentclass[twocolumn,english,aps,prl,floatfix,preprintnumbers,showpacs,amsfonts,amssymb,superscriptaddress]{revtex4}
\usepackage[T1]{fontenc}
\usepackage[latin1]{inputenc}
\usepackage{amsmath}
\usepackage{graphicx}
\usepackage{amssymb}

\makeatletter

\usepackage{amscd}
\usepackage{bm}

\usepackage{babel}
\makeatother
\begin{document}

\title{Dimerization induced by the RKKY interaction}

\author{J.~C.~Xavier}

\affiliation{Instituto de F\'{\i}sica Gleb Wataghin, Unicamp, C.P. 6165, Campinas
SP 13083-970, Brazil}

\author{R. G. Pereira}

\affiliation{Instituto de F\'{\i}sica Gleb Wataghin, Unicamp, C.P. 6165, Campinas
SP 13083-970, Brazil}

\author{E.~Miranda}

\affiliation{Instituto de F\'{\i}sica Gleb Wataghin, Unicamp, C.P. 6165, Campinas
SP 13083-970, Brazil}

\author{I.~Affleck}

\affiliation{Department of Physics, Boston University, 590 Commonwealth Ave.,
Boston, MA 02215 }

\date{\today{}}

\begin{abstract}
We report the presence of spin dimerization in the ground state of the
one dimensional Kondo lattice model at quarter filling. The emergence
of this new phase of the Kondo lattice can be traced to the form of
the RKKY interaction between the localized moments and provides the
first example of dimerization induced indirectly by itinerant
electrons.  We propose this dimer ordering as the driving mechanism of
the spin-Peierls phase observed in the quasi-one-dimensional organic
compounds (Per)$_{2}$M(mnt)$_{2}$ (M=Pt, Pd). Moreover, this suggests
that a richer phase diagram than the Doniach paradigm may be needed to
accommodate the physics of heavy fermion materials.
\end{abstract}

\maketitle
Our understanding of the physics of heavy fermion materials has been
strongly influenced by the paradigm set by Doniach \cite{Doniach}.
It proposes that the ground state of the Kondo lattice model exhibits,
as a function of the Kondo coupling constant, antiferromagnetic and
paramagnetic phases separated by a quantum critical point. A great
deal of effort, both experimental and theoretical, has been devoted
to the elucidation of the nature of this quantum phase transition
(for an overview, see \cite{pierspepinsirevaz}). On the other hand,
the variety of behavior observed in real materials suggests that the
complete picture is far richer.

We illustrate this richness by identifying a new phase of the
one-dimensional Kondo lattice model \cite{Tsunetsugu}. Our numerical
calculations with the density matrix renormalization group (DMRG)
technique show that the ground state of the model has spin
dimerization at quarter filling ($n=1/2)$. We explain how this
previously overlooked state can be understood in terms of the
effective spin-spin RKKY \cite{rkky} interaction mediated by the
conduction electrons.  Interestingly, the quasi-one-dimensional
organic compounds (Per)$_{2}$M(mnt)$_{2}$ (M=Pt, Pd) are \emph{actual
realizations of weakly coupled quarter-filled $S=\frac{1}{2}$ Kondo
chains}
\cite{henriquesetal1,henriquesetal2,bourbonnaisetal,matosetal}.  We
will show, based on our results, that the itinerant-electron-induced
dimer order we have found is a viable candidate for the driving
mechanism behind the hitherto unexplained dimerization transition
observed in these systems.

We consider the one-dimensional $S=\frac{1}{2}$ Kondo lattice
Hamiltonian with $L$ sites \begin{equation} H=-\sum _{j=1,\sigma
}^{L-1}\left(c_{j,\sigma }^{\dagger }c_{j+1,\sigma }^{\phantom \dagger
}+\mathrm{H.c.}\right)+J\sum _{j=1}^{L}\mathbf{S}_{j}\cdot
\mathbf{s}_{j}\label{eq:ham}\end{equation} where $c_{j\sigma }$
annihilates a conduction electron in site $j$ with spin projection
$\sigma $, $\mathbf{S}_{j}$ is a localized spin-$\frac{1}{2}$ operator
and $\mathbf{s}_{j}=\frac{1}{2}\sum _{\alpha \beta }c_{j,\alpha
}^{\dagger }\bm \sigma _{\alpha \beta }c_{j,\beta }^{\phantom \dagger
}$ is the conduction electron spin density operator. $J>0$ is the
Kondo coupling constant between the conduction electrons and the local
moments.  We have set the hopping amplitude and the lattice spacing to
unity to fix the energy and length scales. We treated the model with
the DMRG technique \cite{white,white2} with open boundary conditions.
We used the finite-size algorithm for sizes up to $L=120$ keeping up
to $m=800$ states per block. The discarded weight was typically about
$10^{-5}-10^{-8}$ in the final sweep.

\begin{figure}[htbp]
\begin{center}\includegraphics[  width=3in,
  keepaspectratio]{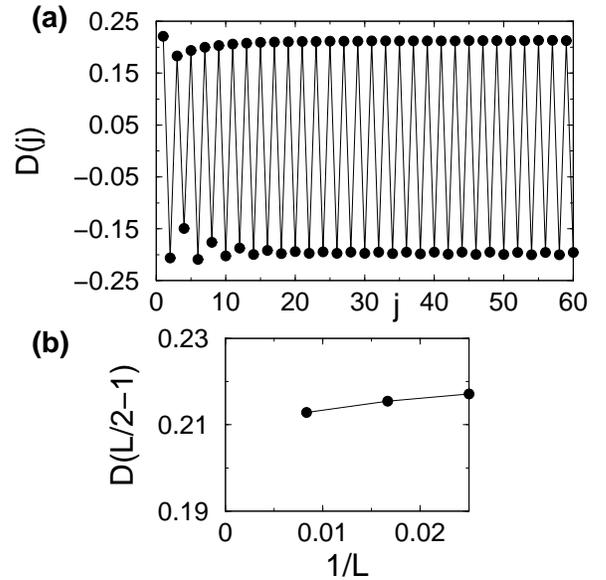}\end{center}
\caption{\label{fig1} (a) The dimer order parameter $D(j)$ vs lattice site
for $J=0.5$, $L=120$ and density $n=0.5$. Only half the chain is
shown. (b) $D(L/2-1)$ vs $1/L$ for $J=0.5$ and $n=0.5$.}
\end{figure}

Dimerization of the localized spins may be detected through the order
parameter $D\left(j\right)=\left\langle \mathbf{S}_{j}\cdot
\mathbf{S}_{j+1}\right\rangle $.  In a uniform system $D(j)$ is
$j$-independent, whereas it shows oscillations of period 2 in the
presence of dimerization. In Fig.~\ref{fig1} (a) we show $D(j)$ at
quarter filling ($n=1/2$) for $J=0.5$ and $L=120$. There are strong
oscillations with amplitude $\approx 0.21$ around an average zero
value. We have checked that this is not an artifact of a finite system
or open boundaries. Indeed, in Fig.~\ref{fig1}~(b), we show the value
of the order parameter at the center of the chain $D(j=L/2-1)$ as a
function of $1/L$. There is very weak size dependence indicating that
the value at $L=120$ is already very close to the thermodynamic
limit. This robust result is unambiguous evidence for a dimerized
ground state at quarter-filling and it is surprising that it has gone
unnoticed in such a well studied model \cite{Tsunetsugu}.  As far as
we know, it is the first example of dimerization induced by an
\emph{indirect} spin-spin interaction mediated by itinerant electrons.

\begin{figure}[htbp]
\begin{center}\includegraphics[  width=3in,
  keepaspectratio]{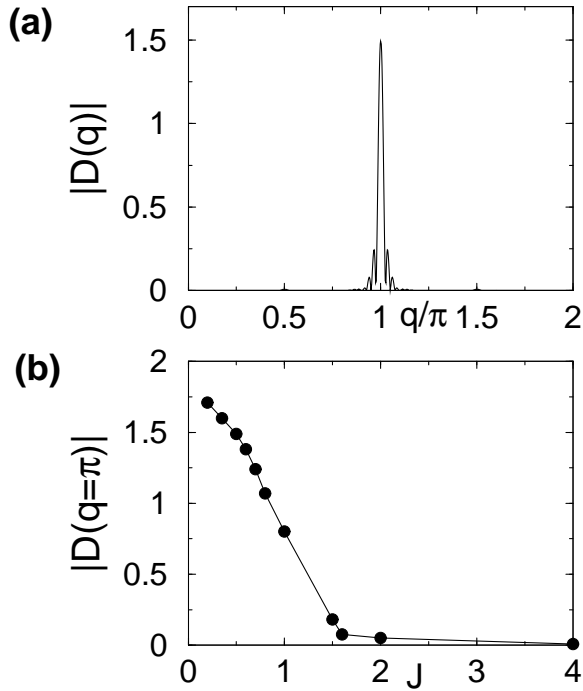}\end{center}
\caption{\label{fig2}(a) Modulus of the smoothed Fourier transform
$\left|D(q)\right|$ vs momentum for $J=0.5$, $L=120$ and density
$n=0.5$. (b) $\left|D(q=\pi )\right|$ as a function of the Kondo
coupling $J$ for $L=120$ and $n=0.5$.}
\end{figure}

It is also instructive to study the Fourier transform of $D(j)$.  In
order to minimize boundary effects we used a smoothed Fourier
transform as suggested in Refs. \cite{vekicwhite,whiteaffscal}.  In
Fig.~\ref{fig2}~(a) we show the smoothed Fourier transform of $D(j)$
for $J=0.5$, $n=0.5$ and $L=120$. The only feature is a noticeable
peak at $q=\pi $ reflecting the alternating sign of the order
parameter $D\left(j\right)$. The critical value of $J$ that separates
the paramagnetic phase from the ferromagnetic one is $J_{c}\sim 1.7$
for $n=0.5$ \cite{Tsunetsugu}. We found that the dimerization occurs
only inside the paramagnetic phase. In Fig.~\ref{fig2} (b) we show the
intensity of the peak of $D(q)$ at $q=\pi $ as a function of the Kondo
coupling $J$ for $L=120$ and $n=0.5$. As can be seen, the dimerization
decreases smoothly as we increase $J$ and disappears in the
ferromagnetic phase (within the accuracy of the DMRG).

\begin{figure}[htbp]
\begin{center}\includegraphics[  width=3in]{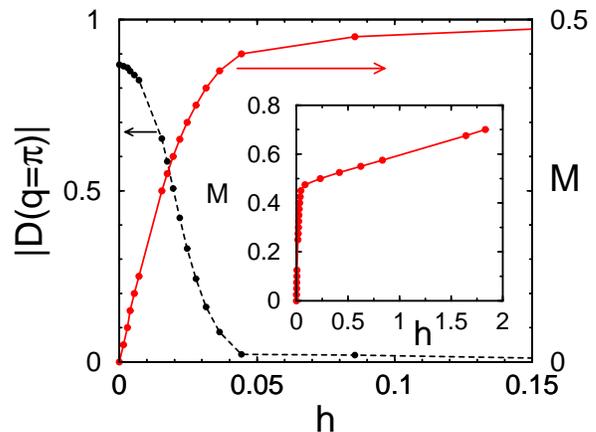}\end{center}
\caption{\label{fig3}Magnetization density $M$ and modulus of the smoothed
Fourier transform $\left|D(q=\pi )\right|$ as a function of applied
magnetic field h, for $L=40$, $J=0.5$ and $n=0.5$. Inset: Magnetization
density up to higher fields.}
\end{figure}

A very small magnetic field is enough to kill the dimerization. We
show in Fig.~\ref{fig3} the magnetization density and the smoothed
Fourier transform of the dimer order parameter as a function of magnetic
field $h$ for $J=0.5$ and $n=0.5$ (making $g\mu _{B}=1$ for both
localized spins and conduction electrons). Since finite-size effects
are negligible, the results here are for $L=40$. There is a characteristic
field $h^{*}\approx 0.05,$ which separates a high susceptibility
region at $h\lesssim h^{*}$ from a low susceptibility one at $h\gtrsim h^{*}$
(see inset of Fig.~\ref{fig3}). Similar results were observed for
other densities and Kondo couplings. As $h$ is increased up to $h^{*}$,
the magnetization density grows linearly up to $M\approx 0.5.$ This
corresponds to the full polarization of the localized spins. A further
increase of $h$ past $h^{*}$ acts to slowly polarize the conduction
electrons (eventually reaching full saturation at $M=0.75$),
which are now effectively decoupled from the local moments. This is
confirmed by the value of the slope for $h\gtrsim h^{*}$, which is
very close to the spin susceptibility of the non-interacting one-dimensional
electron gas. Thus, the field $h^{*}$ is the characteristic energy
scale of the interaction between the spin lattice and the electron
sea. The numerical values of $h^{*}$ and the ratio of high-field
and low-field susceptibilities ($\approx 0.008$) both suggest the
presence of an exponentially small scale, as seen in other studies
\cite{Tsunetsugu}. This scale also governs the suppression of the
dimerization as can be seen in Fig.~\ref{fig3}. Since at $h\gtrsim h^{*}$
the conduction electrons effectively decouple from the localized spins,
this is direct evidence that the dimerization is induced by the interaction
with the itinerant electrons. 

\begin{figure}[htbp]
\begin{center}\includegraphics[  width=2.5in,
  keepaspectratio]{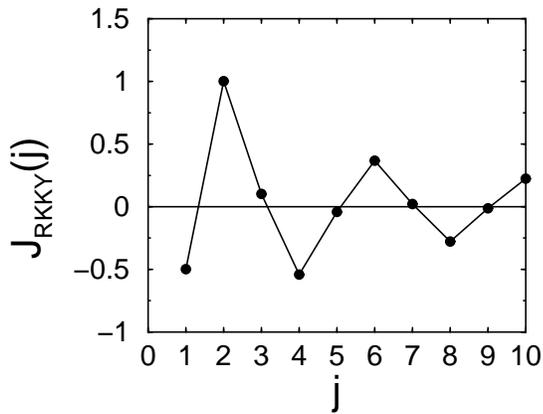}\end{center}
\caption{\label{fig4}RKKY coupling constant (in arbitrary units) as a function
of the distance between sites.}
\end{figure}

Dimerization at quarter-filling is perhaps not too surprising if we
consider the RKKY interaction between localized spins a distance $j$
apart. This is appropriate if $J\ll1$. Using the usual RKKY formula
for a one-dimensional tight-binding lattice at quarter filling we can
get the first ten couplings $J_{RKKY}\left(j\right)$ as shown in
Fig.~\ref{fig4}.  Focusing initially on the first two neighbors, we
see that $J_{RKKY}\left(1\right)<0$ and $J_{RKKY}\left(2\right)>0$.
It is believed that a spin chain with first and second neighbor
interactions $J_1$ and $J_2$ is spontaneously dimerized if $J_2>0$ and
$-4J_2<J_1<0$ \cite{itoihald} and this is satisfied by the RKKY
couplings. In fact, even if $J_1>0$ dimerization would ensue provided
that $J_1\alt 4.15 J_2$ \cite{haldanedimer,majumdarghosh}.

Another curious aspect of Fig.~\ref{fig1} is the sign alternation of
$D(j)$ with an almost constant amplitude $\sim 0.21$. A classical
localized spin configuration of the type $\uparrow\uparrow
\downarrow\downarrow\uparrow\uparrow\cdots$
would give rise to such an alternation with amplitude
$0.25$. Interestingly, the further neighbor RKKY couplings of
Fig.~\ref{fig4} favor this classical state (note that
$J_{RKKY}\left(j\right)\approx 0$ for odd $j\geq 3$). Of course, we do
not expect this long-range classical order to survive in the singlet
ground state, but the fact that $|D(j)|\sim 0.25$ suggests that it may
have a very long correlation length.

\begin{figure}[htbp]
\begin{center}\includegraphics[width=2.5in,
  keepaspectratio]{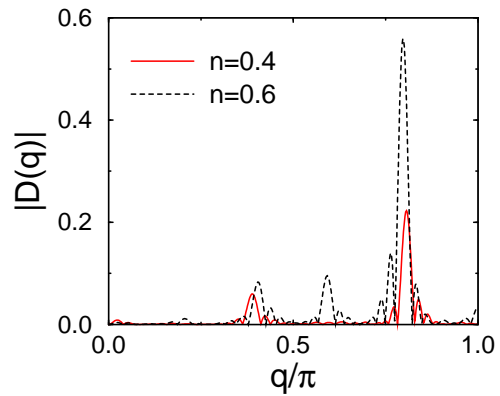}\end{center}
\caption{\label{fig5} Modulus of the smoothed Fourier transform $D(q)$
away from quarter filling ($L=120$ and $J=0.35$). 
}
\end{figure}

We have also calculated dimer correlations away from $n=0.5$. We
observed that $D(j)$ no longer oscillates symmetrically around zero.
The average actually goes from positive (ferromagnetic) for $n<0.5$ to
negative (antiferromagnetic) for $n>0.5$. This is consistent with the
general trend of the RKKY interaction which is predominantly
ferromagnetic at low fillings. However, in contrast to the
quarter-filled case, away from $n=0.5$ $D(j)$ has more than one
characteristic wave vector.  This is shown in Fig.~\ref{fig5} for the
densities $n=0.4$ and $n=0.6$ ($L=120$ and $J=0.35$). They have
$4k_{F}=2\pi n$ (mod $2\pi $) as the dominant peak. There are also
sub-dominant peaks at $2k_{F}^{small}= \pi n$ or $2k_{F}^{large}=\pi
(1-n)$, which are the Fermi sea sizes without and with the localized
moments included in the count, respectively
\cite{shibataetal,yamanakaetal,xavier1}. However, no clear trend can
be discerned.  Quantum fluctuations will destroy incommensurate order
in one dimension. However, the system may lock onto particular
commensurate structures within finite intervals of $n$. Such long
commensurate periods have been discussed in related Kondo lattice
models \cite{shibataishiiueda} used to describe the pnictide CeSb
\cite{rossatmignod} and in the context of doped Mott insulators
\cite{vojtazhangsachdev}.

The organic compounds (Per)$_{2}$M(mnt)$_{2}$, where M=Pt or Pd,
{}``Per'' is perylene and {}``mnt'' is maleonitriledithiolate, are
realizations of quarter-filled quasi-one-dimensional Kondo lattices
\cite{henriquesetal1,bourbonnaisetal,matosetal}. The localized spins
reside in the M(mnt)$_{2}$ units and the conduction electrons are
provided by the perylene groups. These compounds show metal insulator
transitions at $T_{MIT}\left(\mathrm{Pt}\right)=7$ K and
$T_{MIT}\left(\mathrm{Pd}\right)=28$ K, accompanied by one-dimensional
lattice instabilities which indicate the formation of a Peierls-type
dimerization. Initial studies \cite{henriquesetal2} had difficulties
reconciling the behavior of these systems with conventional
electron-phonon (Peierls) or spin-phonon (spin-Peierls) instabilities
and the exact nature of these transitions has remained a mystery.  On
the other hand, there were speculations on the role played by the RKKY
interaction \cite{bourbonnaisetal}. Several uncertainties remained,
however, since spin-density wave (SDW) ordering was not observed and
the Fermi wave vector of the conduction electrons $\left(2k_{F}=\pi
/2\right)$ does not match the period of the dimerization.

\begin{figure}[htbp]
\begin{center}\includegraphics[width=2.5in,
  keepaspectratio]{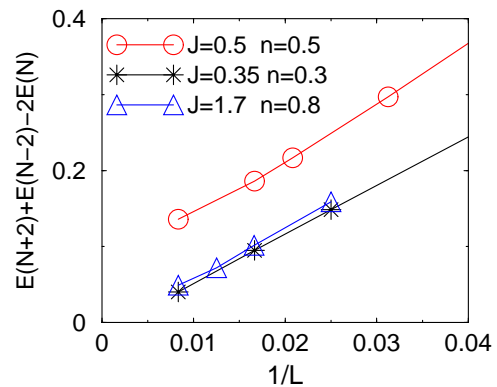}\end{center}
\caption{\label{fig6} Charge gap vs $1/L$
for various fillings.  The notation is the same as in
ref. \cite{whiteaffscal}. The accuracy is given by the size of the
symbols.}
\end{figure}

Our results suggest a possible way out of this quandary. Indeed, the
one-dimensional nature of the system suppresses the appearance of a
conventional SDW and favors a dimerized state instead. Furthermore,
the Fermi wave vector, as explained in connection with
Fig.~\ref{fig4}, is actually crucial for providing the correct spatial
dependence of the RKKY interactions that lead to the dimer order. An
estimate of the finite transition temperature (absent in the strictly
one-dimensional model) could be obtained from the divergence of the
dimer susceptibility as $T\to 0$, along the lines explained in
Refs.~\cite{affleckgelfandsingh,affleckhalperin}.  Our results on the
ground state, however, suggest the presence of a (probably
exponentially) small energy scale of a size compatible with the
transition temperatures (Fig.~\ref{fig3}). It is therefore tempting to
use this scale as an estimate of $T_{MIT}$. Moreover, the direct
measurement of the magnetic field needed to destroy the dimer order at
$T=0$ showed it to be of the order of $k_{B}T_{MIT}/\mu _{B}$
\cite{matosetal}, lending further support to this identification.

Finally, in Fig.~\ref{fig6} we plot the finite size charge gap for
several densities (in the sector of total spin zero) as a function of
$1/L$. It is clear that the quarter-filled case has a charge gap,
unlike other fillings (see Ref. \cite{whiteaffscal} for a similar
case). A mechanism for the opening of this gap can be obtained by
bosonization. By analogy with the RKKY interaction, if we integrate
out the localized spins an effective interaction between conduction
electron spins at sites $i$ and $j$ is generated, which is
proportional to $\left\langle \mathbf{S}_{i}\cdot
\mathbf{S}_{j}\right\rangle .$ The strong dimer correlations of
Fig.~\ref{fig1} lead to a term $\propto
\delta_{i,j+1}\left(-1\right)^{j}$.  This gives rise to a
\emph{staggered} nearest neighbor interaction between conduction
electron spins $\left(-1\right)^j \mathbf{s}_j
\cdot\mathbf{s}_{j+1}$. Bosonizing this \emph{at quarter-filling} leads to
a $\cos \left(\sqrt{8K_\rho}\phi _{\rho}\right)$ term, where we follow
the notation of Ref.~\cite{voit}. If $K_\rho<1$ this term is relevant
and opens a gap in the charge sector in close analogy to the umklapp
term in the half-filled Hubbard model \cite{voit}. This is a very mild
condition and is likely to be fulfilled. Indeed, numerical estimates
of $K_\rho$ at other fillings indicate that $K_\rho<1$
\cite{shibataetal}.  Thus, we propose this RKKY-induced dimerization
as the driving mechanism behind the metal-insulator transitions
observed in (Per)$_{2}$M(mnt)$_{2}$. Note that a similar term is
generated in the spin sector $\cos \left(\sqrt{8K_\sigma}\phi
_{\sigma}\right)$. However, $K_\sigma =1$ due to SU(2) symmetry and
this is a \emph{marginal} term. Its relevance and the accompanying spin
gap depends on the sign of the coupling constant. We have found no
evidence of a spin gap at this filling.

In conclusion, we have found that the Kondo lattice model in one
dimension has a novel type of dimerization at quarter conduction
electron filling. It can be understood from the structure of the RKKY
interaction between localized moments.  Furthermore, it provides a
simple mechanism for the metal-insulator transition of some
quasi-one-dimensional organic compounds. Although confined to one
dimension, we believe our results may have implications for the phase
diagram of three-dimensional heavy fermion materials.

We thank A. L. Malvezzi for providing some DMRG data for comparison
and A. Villares Ferrer for useful discussions.  This work was
supported by FAPESP 00/02802-7 (JCX), 01/12160-5 (RGP) and 01/00719-8
(JCX, EM), CNPq 301222/97-5 (EM), and NSF DMR-0203159 (IA).

\end{document}